\documentclass[12pt]{iopart}

\usepackage{graphicx}
\usepackage{dcolumn}
\usepackage{caption}

\begin{document}
\title[Two-particle angular correlations in p+p, Cu+Cu and Au+Au]{System size dependence of two-particle angular correlations in p+p, Cu+Cu and Au+Au collisions}
\author{Wei Li for the PHOBOS collaboration}
\address{
\vspace{2mm}
{\scriptsize
Massachusetts Institute of Technology, 77 Mass Ave, Cambridge, MA 02139-4307, USA\\
}}
\ead{davidlw@mit.edu}
{\footnotesize
B.Alver$^4$,
B.B.Back$^1$,
M.D.Baker$^2$,
M.Ballintijn$^4$,
D.S.Barton$^2$,
R.R.Betts$^6$,
A.A.Bickley$^7$,
R.Bindel$^7$,
W.Busza$^4$,
A.Carroll$^2$,
Z.Chai$^2$,
V.Chetluru$^6$,
M.P.Decowski$^4$,
E.Garc\'{\i}a$^6$,
N.George$^2$,
T.Gburek$^3$,
K.Gulbrandsen$^4$,
C.Halliwell$^6$,
J.Hamblen$^8$,
I.Harnarine$^6$,
M.Hauer$^2$,
C.Henderson$^4$,
D.J.Hofman$^6$,
R.S.Hollis$^6$,
R.Ho\l y\'{n}ski$^3$,
B.Holzman$^2$,
A.Iordanova$^6$,
E.Johnson$^8$,
J.L.Kane$^4$,
N.Khan$^8$,
P.Kulinich$^4$,
C.M.Kuo$^5$,
W.Li$^4$,
W.T.Lin$^5$,
C.Loizides$^4$,
S.Manly$^8$,
A.C.Mignerey$^7$,
R.Nouicer$^2$,
A.Olszewski$^3$,
R.Pak$^2$,
C.Reed$^4$,
E.Richardson$^7$,
C.Roland$^4$,
G.Roland$^4$,
J.Sagerer$^6$,
H.Seals$^2$,
I.Sedykh$^2$,
C.E.Smith$^6$,
M.A.Stankiewicz$^2$,
P.Steinberg$^2$,
G.S.F.Stephans$^4$,
A.Sukhanov$^2$,
A.Szostak$^2$,
M.B.Tonjes$^7$,
A.Trzupek$^3$,
C.Vale$^4$,
G.J.van~Nieuwenhuizen$^4$,
S.S.Vaurynovich$^4$,
R.Verdier$^4$,
G.I.Veres$^4$,
P.Walters$^8$,
E.Wenger$^4$,
D.Willhelm$^7$,
F.L.H.Wolfs$^8$,
B.Wosiek$^3$,
K.Wo\'{z}niak$^3$,
S.Wyngaardt$^2$,
B.Wys\l ouch$^4$\\
}
\vspace{3mm}
\small

{\scriptsize
\noindent
$^1$~Argonne National Laboratory, Argonne, IL 60439-4843, USA\\
$^2$~Brookhaven National Laboratory, Upton, NY 11973-5000, USA\\
$^3$~Institute of Nuclear Physics PAN, Krak\'{o}w, Poland\\
$^4$~Massachusetts Institute of Technology, Cambridge, MA 02139-4307, USA\\
$^5$~National Central University, Chung-Li, Taiwan\\
$^6$~University of Illinois at Chicago, Chicago, IL 60607-7059, USA\\
$^7$~University of Maryland, College Park, MD 20742, USA\\
$^8$~University of Rochester, Rochester, NY 14627, USA\\
}

\begin{abstract}

We present a systematic study of two-particle angular correlations in p+p, Cu+Cu and Au+Au
collisions over a broad range of pseudorapidity and azimuthal angle. The PHOBOS detector has a uniquely large 
angular coverage for inclusive charged particles, which allows for the study of correlations 
on both long- and short-range scales. A complex two-dimensional correlation structure 
emerges which is interpreted in the context of a cluster model. The cluster size and its 
decay width are extracted from the two-particle pseudorapidity correlation function. 
The cluster size found in semi-central Cu+Cu and Au+Au collisions is comparable to that 
found in p+p but a non-trivial increase of cluster size with decreasing centrality is observed.
Moreover, the comparison between Cu+Cu and Au+Au 
systems shows an interesting scaling of the cluster size with the measured fraction of total 
cross section (which is related to $b/2R$), suggesting a geometric origin. These results should provide 
insight into the hadronization stage of the hot and dense medium created in heavy ion collisions. 
\end{abstract}


\vspace{-0.4cm}

Multiparticle correlation analyses have proven to be a powerful tool in exploring the 
underlying mechanism of particle production in high energy hadronic collisions. Both short- 
and long-range correlations have been discovered in the past decades~\cite{UA5_3energy,ISR_twolowenergy,phobos_pp} 
suggesting that particles tend to be produced in a correlated fashion~\cite{cluster_model,cluster_fit}.
In this scenario, hadronization proceeds via ``clusters'', high mass states which decay isotropically 
in their center of mass frame to final-state hadrons. Two-particle angular 
correlations can provide detailed information about the 
cluster properties, e.g. their multiplicity (``size'') and extent in phase space (``width''). 
In heavy ion collisions at RHIC, the expected formation of a Quark Gluon Plasma (QGP) could 
lead to a modification of the clusters relative to p+p collisions~\cite{AAcluster_prediction}. 
A comprehensive analysis of cluster properties in p+p and A+A collisions should provide essential information 
for understanding the hadronization stage in A+A collisions.


Covering pseudorapidity range ($\eta=-\ln(\tan(\theta/2))$) $-3<\eta<3$ over almost full 
azimuthal angle, the PHOBOS Octagon detector~\cite{phobos_detector} is ideally suited for 
studying the angular correlations of the particles emitted from clusters. 
The detailed analysis procedure has been described in Ref.~\cite{phobos_pp}. The inclusive two-particle 
correlation function in ($\Delta \eta,\Delta \phi$) space is defined as follows:

\begin{equation}
\label{2pcorr_incl}
R(\Delta \eta,\Delta \phi)=\left\langle(n-1)\left(\frac{F_{n}
(\Delta \eta,\Delta \phi)}{B_{n}(\Delta \eta,\Delta \phi)}-1\right)\right>
\end{equation} 

\noindent where $F_{n}(\Delta \eta,\Delta \phi)$ is the foreground distribution obtained 
by taking two-particle pairs from the same event and $B_{n}(\Delta \eta,\Delta \phi)$ is the 
background distribution constructed by randomly selecting two particles from two different 
events with similar vertex position and centrality. The event multiplicity, $n$, is introduced 
to compensate for the trivial dilution effects from uncorrelated particles. $R(\Delta \eta,\Delta \phi)$ 
is defined in such a way that if a heavy ion collision is simply a superposition of individual p+p 
collisions, the same correlation function should be observed.


\begin{figure}[t]
\captionsetup[figure]{margin=0.1cm,font=small}

\begin{minipage}[t]{0.32\textwidth}
 \centerline{
  \includegraphics[width=\textwidth]{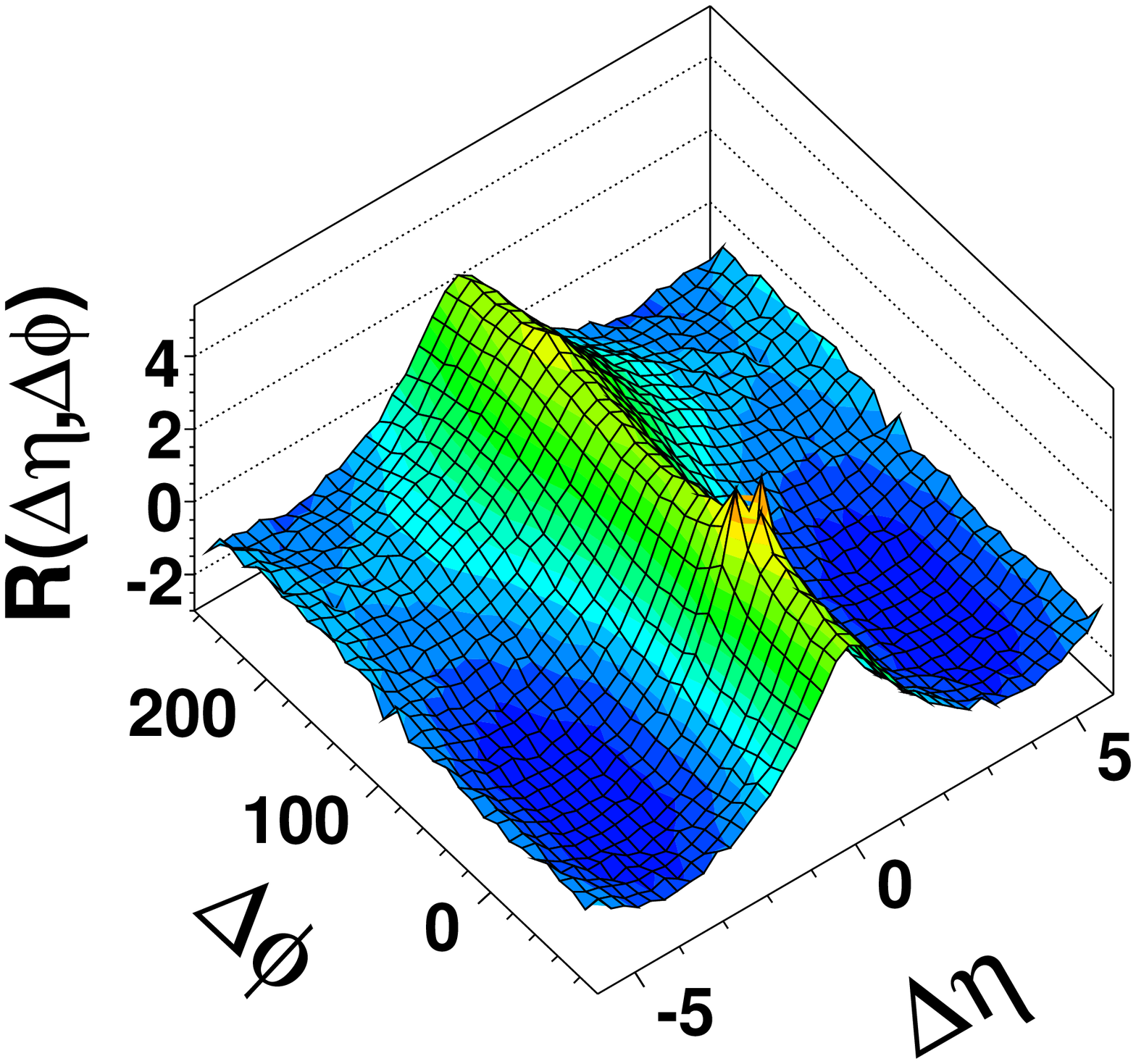}
 }
 \vspace{-0.4cm}
 \caption{Inclusive two-particle correlation 
          function in ($\Delta \eta$, $\Delta \phi$) for 
          p+p collisions at $\sqrt{s}$ = 200~GeV~\cite{phobos_pp}.}
 \label{pp200_2D_corrected}
\end{minipage}
\hspace{\fill}
\begin{minipage}[t]{0.32\textwidth}
 \centerline{
  \includegraphics[width=\textwidth]{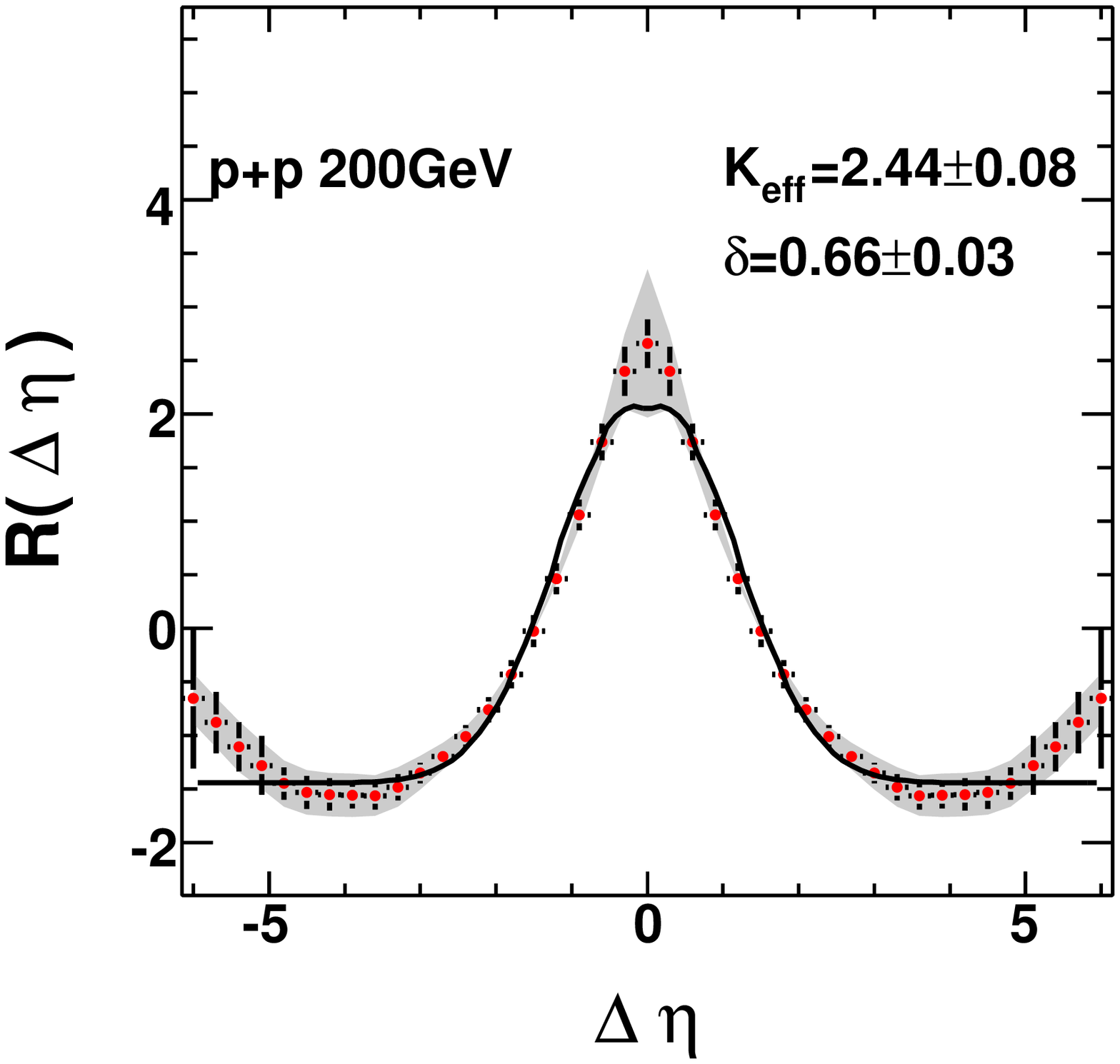}
 }
 \vspace{-0.5cm}
 \caption{1D two-particle pseudorapidity correlation function in 
          $\Delta \eta$ for p+p collisions at $\sqrt{s}$ = 200~GeV 
          together with a fit from a cluster model~\cite{phobos_pp}.}
 \label{pp200_eta_corrected_fit}
\end{minipage}
\hspace{\fill}
\begin{minipage}[t]{0.32\textwidth}
 \centerline{
  \includegraphics[width=\textwidth]{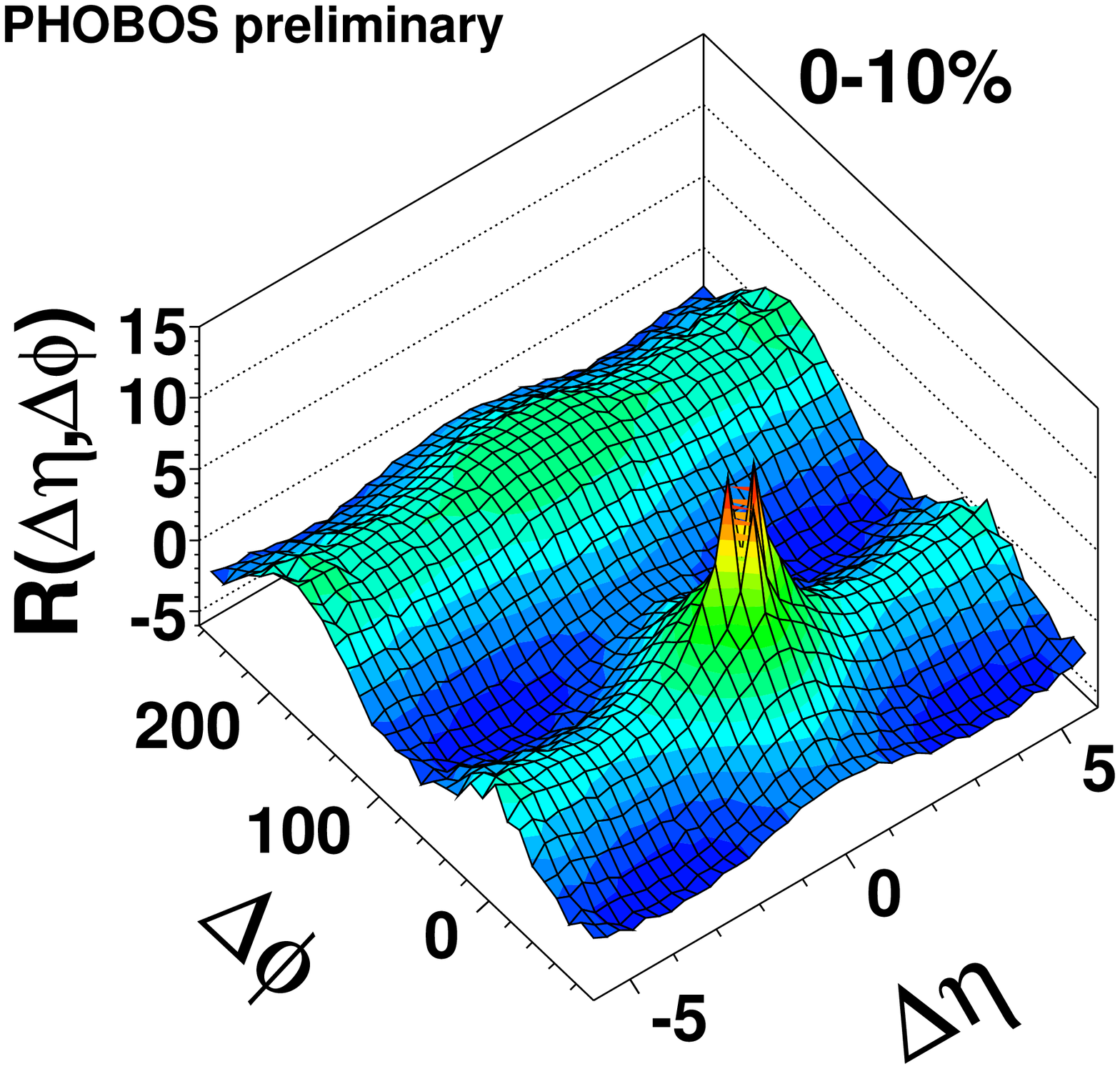}
 }
 \vspace{-0.5cm}
 \caption{Two-particle correlation 
          function in ($\Delta \eta$, $\Delta \phi$) 
          for the most central 10\% of the Au+Au collisions at \mbox{$\sqrt{s_{_{NN}}}$ = 200~GeV}.}
 \label{AuAu_2D_binstart15}
\end{minipage}
\end{figure}

Fig.~\ref{pp200_2D_corrected} shows the two-particle inclusive correlation function in p+p 
collisions at \mbox{$\sqrt{s}$ = 200~GeV} as a function of $\Delta \eta$ and $\Delta \phi$. 
A set of correction procedures has been applied, based on MC simulations, to extract the
correlations between primary particles. The complex correlation structure suggests that the 
short range correlation is approximately Gaussian in $\Delta \eta$ and persists over the full 
$\Delta \phi$ range, becoming broader toward higher $\Delta \phi$. If clusters are the precursors 
to the final measured hadrons, a high $p_{T}$ cluster would contribute to a narrow peak at 
the near-side ($\Delta \phi$ near $0^{o}$) region of the correlation function in Fig.~\ref{pp200_2D_corrected}, 
while a lower $p_{T}$ cluster will contribute to the broader away-side.

To study one aspect of the correlation function quantitatively, 
the two-dimensional (2D) correlation function is 
projected into a one-dimensional (1D) correlation function $R(\Delta \eta)$ shown in 
Fig.~\ref{2pcorr_clusterfitting_incl}. It is fit to a functional form derived in Ref.~\cite{cluster_fit} 
in an independent cluster emission model:

\begin{equation}
\label{2pcorr_clusterfitting_incl}
R(\Delta \eta)=\alpha\left[\frac{\Gamma(\Delta \eta)}{B(\Delta \eta)}-1\right]   
\end{equation}

\noindent where $B(\Delta \eta)$ is the background distribution obtained by event-mixing. 
The parameter $\alpha=\frac{\langle K(K-1) \rangle}{\langle K \rangle}$ contains the information about the cluster 
size $K$ and $\Gamma(\Delta \eta)$ is a Gaussian function $\propto exp{(- (\Delta \eta)^{2}/(4\delta^{2}))}$ 
characterizing the correlation of particles produced by a single cluster, where $\delta$ 
corresponds to the decay width of the clusters in $\eta$ space. The effective cluster multiplicity, 
or ``size'' is defined to be $K_{\rm eff}=\frac{\langle K(K-1) \rangle}{\langle K \rangle}+1=\langle K \rangle+\frac{\sigma_{K}^{2}}{\langle K \rangle}$.
Of course, without any knowledge of the distribution of $K$, it is impossible to directly 
measure the average cluster size $\langle K \rangle$. However, by a $\chi^{2}$ fit of Eq.~\ref{2pcorr_clusterfitting_incl} 
to the measured two-particle correlation function, an example of which is shown in 
Fig.~\ref{pp200_eta_corrected_fit},
the effective cluster size $K_{\rm eff}$ can be estimated, as well as the cluster decay width $\delta$. 
A $K_{\rm eff}$ of about 2.5 indicates that on average every charged particle is produced in 
association with another 1.5 particles, if it is assumed that $\sigma_{K}^{2} \approx 0$. 

\begin{figure}[t]
\captionsetup[figure]{margin=0.1cm,font=small}
\begin{minipage}[t]{0.45\textwidth}
 \centerline{
  \includegraphics[width=\textwidth]{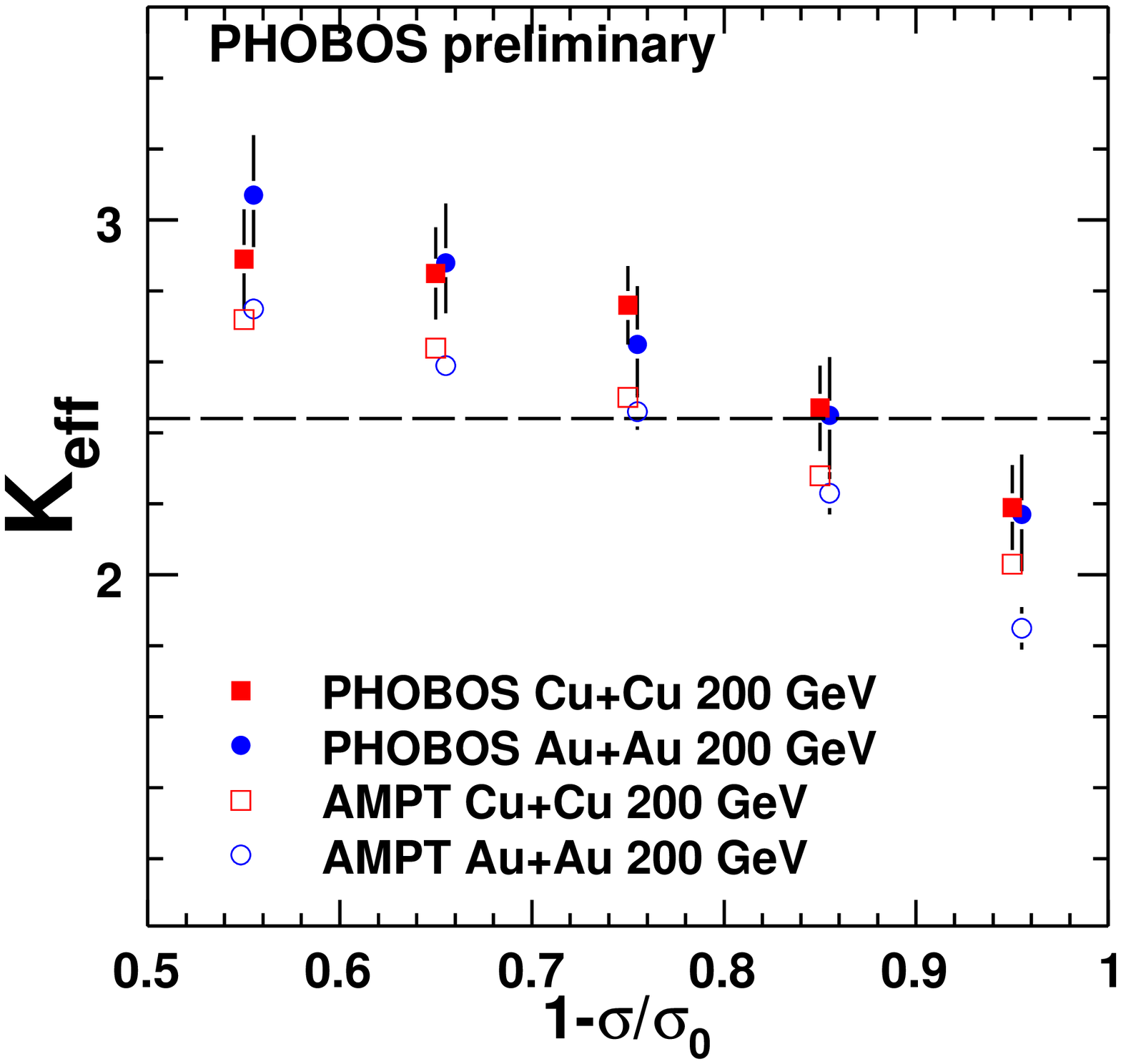}
 }
 \vspace{-0.5cm}
 \caption{$K_{\rm eff}$ as a function of fractional cross section
          measured by PHOBOS (solid symbols) and from the AMPT model 
          (open symbols) in Cu+Cu (squares) and Au+Au (circles) collisions 
          at $\sqrt{s_{_{NN}}}$ = 200~GeV.}
 \label{cluster_k_cross_inclusive}
\end{minipage}
\hspace{\fill}
\begin{minipage}[t]{0.45\textwidth}
 \centerline{
  \includegraphics[width=\textwidth]{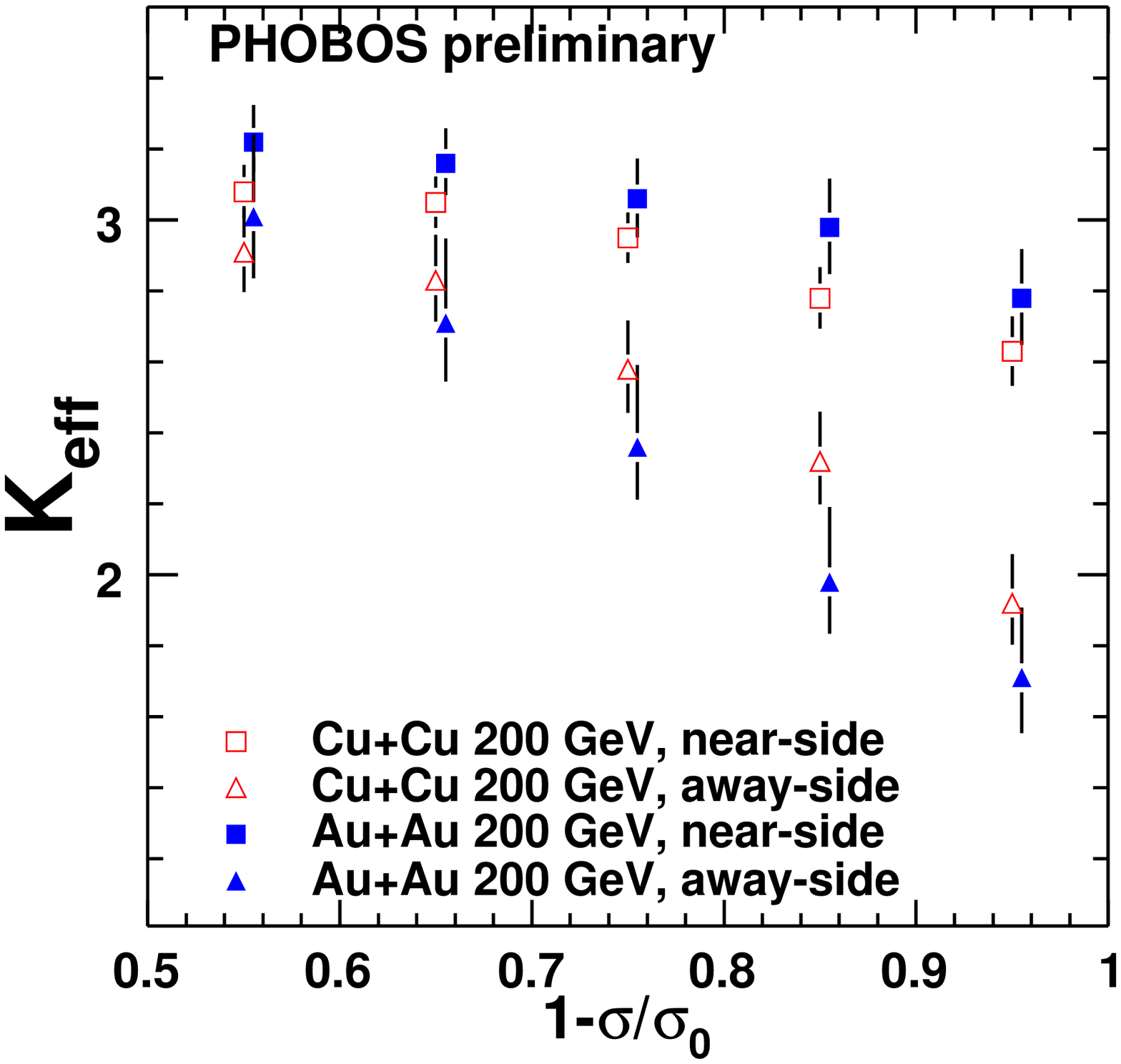}
 }
 \vspace{-0.5cm}
 \caption{Near-side (squares) and away-side (triangles) $K_{\rm eff}$ 
          as a function of fractional cross section measured by PHOBOS
          in Cu+Cu (open symbols) and Au+Au (solid symbols) collisions at $\sqrt{s_{_{NN}}}$ = 200~GeV.}
 \label{cluster_k_cross_NearAway_data}
\end{minipage}
\end{figure}

In heavy ion collisions, not only the cluster-like structure but also
a $\cos(2\Delta\phi)$ elliptic flow component is observed in the 2D correlation 
function as shown in Fig.~\ref{AuAu_2D_binstart15} 
for the most central 10\% of the Au+Au collisions at $\sqrt{s}$ = 200~GeV. 
As for p+p, we average over $\Delta\phi$
in order to study only the cluster properties in pseudorapidity space. 
In this procedure, the flow signal averages to zero.
As described above, the two-particle pseudorapidity correlation function in A+A is fit by 
Eq.~\ref{2pcorr_clusterfitting_incl} in a similar way to p+p. The resulting effective 
cluster size as a function of fractional cross section (collision centrality) 
is shown in Fig.~\ref{cluster_k_cross_inclusive} for Cu+Cu and Au+Au collisions at $\sqrt{s_{_{NN}}}$ = 200~GeV. 
The dashed line indicates the value found in $\sqrt{s}$ = 200~GeV p+p collisions, which suggests 
that the cluster properties are similar in p+p and A+A systems. This implies that the 
phenomenological properties of hadronization appear to be similar in p+p and A+A systems. However, it is also
observed that the cluster size systematically decreases with increasing collision centrality in both Cu+Cu 
and Au+Au collisions. Furthermore, by comparing the two systems at the same fraction of 
the inelastic cross section (which is related to $b/2R$) it appears that the cluster size 
scales with collision geometry of the system, e.g. the shape of the overlap region. 
This feature is unexpected since the information of clusters is extracted from 
pseudorapidity space and not directly connected to the transverse geometry of the system. 
In comparing the data with dynamical models, AMPT gives the same qualitative trend as the data. 
Note that the values of $K_{\rm eff}$ are extracted in a limited acceptance of $|\eta|<3$, and therefore
are slightly smaller than for a full acceptance measurement.
Detailed studies are underway in order to quantify this acceptance effect.  
The decrease in cluster size with centrality in AMPT is related to  
the hadronic rescattering stage. Turning off hadronic processes leads to a larger 
cluster size in both Au+Au and Cu+Cu that is approximately invariant for all centralities. 

Further detailed studies on cluster properties have also been performed. Instead of averaging over
the whole $\Delta\phi$ region, the near- and away-side cluster size can be extracted in a limited 
$\Delta\phi$ range of ($0^{\circ}$,$90^{\circ}$) and ($90^{\circ}$,$180^{\circ}$) respectively. 
In this restricted averaging, the $\cos(2\Delta\phi)$ elliptic flow component again averages to zero.
The results are shown in Fig.~\ref{cluster_k_cross_NearAway_data} as a function of fractional cross section
for Cu+Cu and Au+Au collisions at $\sqrt{s_{_{NN}}}$ = 200~GeV. For the more central collisions,
the away-side cluster size decreases by about 30-40\%, whereas the near-side cluster size decreases more 
slowly. Such a behavior could be understood in a scenario if the medium is extremely dense and 
only clusters produced close to the surface can survive. Then, for away-side clusters, it is more 
likely that part of its decay particles travel into the medium and get absorbed, resulting in 
the loss of away-side correlations. As for the observed collision geometry scaling of the cluster size, 
it might be related to the surface to volume ratio of the system. More detailed modeling is still 
being investigated to understand these phenomena.


In conclusion, the two-particle correlation function for inclusive charged particles 
has been extensively studied over a broad range in $\Delta\eta$ and $\Delta\phi$ 
in p+p, Cu+Cu and Au+Au collisions. 
In particular, it has already been shown that the observed short-range correlations in pseudorapidity have 
a natural interpretation in terms of clusters. In this approach, multiple particles are understood 
to be emitted close together in phase space, with a typical cluster size of 2-3 in p+p collisions.
In the new A+A data, clusters have a similar size but show a non-trivial decrease in size with 
increasing centrality and a geometry scaling feature between Cu+Cu and Au+Au reactions. Analysis
of near- and away-side clusters provides additional information on the details of the cluster properties.
Future work will focus on model studies in order to understand the observed phenomena
in heavy ion collisions.
\vspace{-0.5cm}

\section*{Acknowledgments}

%
%
%
%
This work was partially supported by U.S. DOE grants 
DE-AC02-98CH10886,
DE-FG02-93ER40802, 
DE-FG02-94ER40818,  
DE-FG02-94ER40865, 
DE-FG02-99ER41099, and
DE-AC02-06CH11357, by U.S. 
NSF grants 9603486, 
0072204,            
and 0245011,        
by Polish MNiSW grant N N202 282234 (2008-2010),
by NSC of Taiwan Contract NSC 89-2112-M-008-024, and
by Hungarian OTKA grant (F 049823).


\section*{References}

\begin{thebibliography}{99}
\vspace{-0.5cm}
\bibitem{UA5_3energy} Ansorge\ R\ E \textit{et al.} [UA5] 1988 {\it Z.\ Phys. - Particle and Fields} C \textbf{37} 191  
\bibitem{ISR_twolowenergy} Eggert\ K\ \textit{et al.} 1975 {\it Nucl. Phys.} B \textbf{86} 201  
\bibitem{phobos_pp} B.~Alver \textit{et al.} [PHOBOS] 2007 {\it Phys. Rev.} C \textbf{75} 054913 
\bibitem{cluster_model} Berger\ E\ L\ 1975 {\it Nucl. Phys.} B \textbf{85} 61
\bibitem{cluster_fit} Morel\ A\ and Plaut\ G\ 1974 {\it Nucl. Phys.} B \textbf{78} 541  
\bibitem{AAcluster_prediction} Shi\ L\ J and Jeon\ S 2003 {\it Phys. Rev.} C \textbf{72} 034904 
\bibitem{phobos_detector}  Back\ B\ B \textit{et al.} [PHOBOS] 2003 {\it Nucl.Inst.Meth} A \textbf{499} 603  

\end{thebibliography}
\end{document}